\documentclass[10pt,journal]{IEEEtran}

\usepackage[dvips]{graphicx}
\graphicspath{{./figures/}}
\DeclareGraphicsExtensions{.eps}
\usepackage{float}

\usepackage[cmex10]{amsmath}
\usepackage{amsfonts}
\usepackage{mathtools}
\usepackage{amssymb}

\usepackage{algorithmic}
\usepackage{algorithm}
\algsetup{linenosize=\scriptsize}
\makeatletter
\newcommand\fs@norules{\def\@fs@cfont{\bfseries}\let\@fs@capt\floatc@ruled
  \def\@fs@pre{}%
  \def\@fs@post{}%
  \def\@fs@mid{\kern3pt}%
  \let\@fs@iftopcapt\iftrue}
\makeatother
\floatstyle{norules}
\restylefloat{algorithm}

\usepackage[tight,footnotesize]{subfigure}
\usepackage[font=small,skip=5pt]{caption}
\usepackage{wrapfig}
\usepackage{lipsum}

\usepackage[usenames, dvipsnames]{color}

\usepackage{times} 

\renewcommand{\labelenumii}{\theenumii}
\renewcommand{\theenumii}{\theenumi.\arabic{enumii}.}
\usepackage{enumitem}
\setitemize{noitemsep,topsep=0pt,parsep=0pt,partopsep=0pt}

\newcommand{\protocol}{$\operatorname{Flock}$}

\newtheorem{theorem}{Theorem}[section]
\newtheorem{lemma}[theorem]{Lemma}

\newenvironment{proof}[1][Proof]{\begin{trivlist}
\item[\hskip \labelsep {\bfseries #1}]}{\end{trivlist}}

\newcommand{\qed}{\nobreak \ifvmode \relax \else
      \ifdim\lastskip<1.5em \hskip-\lastskip
      \hskip1.5em plus0em minus0.5em \fi \nobreak
      \vrule height0.75em width0.5em depth0.25em\fi}

\newcommand{\comment}[1]{ }

\usepackage{filecontents}

\usepackage[printonlyused]{acronym}
\acrodef{vm}[VM]{Virtual Machine}
\acrodef{iot}[IoT]{Internet of Things}
\acrodef{rtt}[RTT]{Round Trip Time}
\acrodef{ec}[EC]{Edge Cloud}
\acrodef{ne}[NE]{Nash Equilibrium}
\acrodef{mdp}[MDP]{Markov Decision Process}
\acrodef{poa}[PoA]{Price of Anarchy}

\begin{document}




\title{Flocking Virtual Machines in Quest for Responsive IoT Cloud Services}

\author{Sherif~Abdelwahab and Bechir Hamdaoui
~\\
\small School of EECS, Oregon State University~\\ \{abdelwas,hamdaoui\}@eecs.oregonstate.edu\\
}

\maketitle
\pagenumbering{gobble}

\begin{abstract}
%
We propose \protocol; a simple and scalable protocol that enables live migration of \acp{vm} across heterogeneous edge and conventional cloud platforms to improve the responsiveness of cloud services. \protocol~is designed with properties that are suitable for the use cases of the \acl{iot} (IoT).
%
%
We describe the properties of regularized latency measurements that \protocol~can use for asynchronous and autonomous migration decisions. Such decisions allow communicating \acp{vm} to follow a \emph{flocking-like} behavior that consists of three simple rules: separation, alignment, and cohesion.
Using game theory, we derive analytical bounds on \protocol's \ac{poa}, and prove that flocking \acp{vm} converge to a \acl{ne} while settling in the best possible cloud platforms.
%
We verify the effectiveness of \protocol~through simulations and discuss how its generic objective can simply be tweaked to achieve other objectives, such as cloud load balancing and energy consumption minimization.
\end{abstract}


\IEEEpeerreviewmaketitle

\section{Introduction}
\label{intro}
 \acresetall

Live \ac{vm} migration is essential for improving the responsiveness of cloud services.
\acp{vm}, hosted in \ac{ec} platforms, install resource-rich cloud services close to data sources and users to realize \ac{iot} applications.
Such \acp{vm} provide, for example, real-time video analytics services from cameras to mobile applications.
Hosting \acp{vm} near the edge can subdue the end-to-end services latency from hundreds to tens of milliseconds compared to hosting \acp{vm} in conventional cloud platforms \cite{li2010cloudcmp,satyanarayanan2014cloudlets}.
Existing resource provisioning algorithms migrate \acp{vm} from one \ac{ec} to the other in response to device mobility and service computational capacity requirements. \ac{vm} migration ensures minimal average latency between mobile devices and the \ac{ec} \cite{urgaonkar2015dynamic, jiaoptimal}.
Migration can achieve other goals such as load balancing, efficient service chaining and orchestration, infrastructure cost minimization, efficient content delivery, and energy efficiency, to name a few \cite{chen2014distributed}.

However, migrating \acp{vm} in response to the mobility of \ac{iot} devices is a limited decision given the diverse \ac{iot} use cases.
Such a mobility-triggered migration assumes a constraining \ac{iot} use case in which a cloud service, installed in \acp{vm}, communicates with one - and only one - \ac{iot} device \cite{satyanarayanan2014cloudlets}. Consider \ac{ec} services for smart glasses as an example.
Migrating a \ac{vm}, which hosts video processing cloud services of a Google glass, minimizes the video processing latency as the glass/user moves,
a goal that is reasonable to target only for the Google Glass use case, in which the Google Glass is the singleton client that uses the \ac{ec} video processing services \cite{satyanarayanan2014cloudlets, urgaonkar2015dynamic}.
However, in general \ac{iot} use cases, several \acp{vm}, very likely to be deployed in different \acp{ec}, are needed to execute distributed algorithms (e.g.  computer vision feature extraction, consensus, and aggregation \cite{tron2011distributed}) and require intensive communication among the \acp{vm} and/or the devices. Distributed computer vision, for example, enables distributed context-aware applications such as autonomous vehicles and intelligent traffic systems \cite{ding2012collaborative}.
Predominantly for modern \ac{iot} applications, developers implement analytics via large-scale distributed algorithms executed by \acp{vm} in geographically distributed and heterogeneous cloud platforms \cite{hong2013mobile, abdelwahab2015cloud}.

In this paper, we propose \protocol; a simple protocol by which \acp{vm} autonomously migrate between heterogeneous cloud platforms to minimize their weighted latency measured with end-user applications, \ac{iot} devices, and other peer-\acp{vm}.
In \protocol, a \ac{vm} uses only local latency measurements, processing latency of its hosting cloud, and local information provided by its hosting cloud about other clouds which the \ac{vm} can migrate to.
A \ac{vm} greedily migrates to a cloud platform that reduces its regularized weighted latency.
We prove, using game-theory, that \protocol~converges to a \ac{ne} and its \ac{poa} is $(1+\epsilon)$ given the properties of our proposed social value function.
We discuss how \protocol~serves a generic goal that we can redesign using simple tweaks to achieve other objectives such as load balancing and energy minimization.
Our design allows \acp{vm} to imitate a \emph{flocking-like} behavior in birds: separation, alignment, and cohesion.

\comment{
Existing migration solutions are limited in their applicability to minimize the weighted latency of \acp{vm} with their peers.
Several existing solutions rely on a system-wide central controller to manage the states of \acp{vm}, devices mobility, and physical resources of \acp{ec} or conventional clouds \cite{urgaonkar2015dynamic, eramo2014study}.
These solutions lack scalability for an Internet-sized network such as the \ac{iot} without relaxations that potentially compromise solutions quality.

Consider \ac{mdp} based solutions.
\ac{mdp} requires a central server to collect statistics of user mobility, \acp{vm} utilization, and clouds connectivity and utilization.
This server also executes the value iteration algorithm to evaluate the optimal \acp{vm} migration policy \cite{urgaonkar2015dynamic, eramo2014study, chen2014distributed}.
It is intractable to model all possible states of \acp{vm} and their hosting platforms; hence it is common to discretize states measurements to relax the complexity of the policy optimization algorithms \cite{chen2014distributed, urgaonkar2015dynamic}.
This compromises the solutions quality.

Game-theoretic approaches potentially decentralize the migration algorithms and improve their scalability.
However existing game-theoretic solutions provide an unbounded \ac{poa}\cite{xiao2015solution}.
We can not use them - as is - and guarantee optimal or close to optimal \acp{vm} responsiveness;
\emph{Does a solution with a tight \ac{poa} exist?}
It is also unclear if the current game-theoretic solutions reach an equilibrium considering the weighted  latency as an objective that involves peer-to-peer relations among \acp{vm};
\emph{Under which conditions is an equilibrium guaranteed?}

Finally, existing \ac{vm} migration solutions serve specialized cloud providers objectives (e.g. energy, load, and cost) to profitably manage providers' infrastructure \cite{duong2014joint, xiao2015solution}.
\emph{The existing models do not capture network latency between peer-\acp{vm} that are executing distributed \ac{iot} applications.}
%

}


\section{System Model and Objective}
\label{model}

\newcommand{\game}{\Gamma} 
\newcommand{\Nset}{N} 
\newcommand{\Fset}{F}
\renewcommand{\iint}{i} 
\newcommand{\jint}{j} 
\newcommand{\nint}{n} 
\newcommand{\mint}{m} 
\newcommand{\Eset}{E} 
\newcommand{\eint}{e} 
\newcommand{\wfloat}{w} 
\newcommand{\tfloat}{t} 
\newcommand{\lfloat}{l} 
\newcommand{\pfloat}{p} 
\newcommand{\Sset}{S} 
\newcommand{\sint}{s} 
\newcommand{\otuple}{\sigma} 
\newcommand{\xfloat}{x} 
\newcommand{\ffloat}{f} 
\newcommand{\Cfloat}{C} 
\newcommand{\Rset}{\mathbb{R}} 
\newcommand{\dvec}{d} 
\newcommand{\dfloat}{d} 
\newcommand{\afloat}{a} 
\newcommand{\kint}{k} 
\newcommand{\rvec}{\mathrm{r}} 
\newcommand{\tvec}{\mathrm{t}} 
\newcommand{\lvec}{\mathrm{l}} 
\newcommand{\pvec}{\mathrm{p}} 
\newcommand{\onesvec}{\mathrm{1}} 
\newcommand{\rfloat}{r} 
\newcommand{\alphafloat}{\alpha}
\newcommand{\etafloat}{\eta}
\newcommand{\lambdafloat}{\lambda}
\newcommand{\mufloat}{\mu}
\renewcommand{\Fset}{\mathcal{F}}
\newcommand{\Mset}{\mathcal{M}}
\newcommand{\integ}{\mathbb{Z}}
\newcommand{\real}{\mathbb{R}}
\newcommand{\deltafloat}{\delta}
\newcommand{\migrate}[3]{#2 \xrightarrow{#1} #3}

\renewcommand{\labelenumii}{\theenumii}
\renewcommand{\theenumii}{\theenumi.\arabic{enumii}.}

We consider a network of \acp{vm} modeled as a graph $G=(V,P)$, where $V$ denotes the set of $n$ \acp{vm} and $P$ denotes the set of \ac{vm} pairs such that $p=(i,j) \in P$ if the $i$-th and $j$-th \acp{vm} communicate with each other.
Let $d_{ij} \in \Rset^+$ denote the traffic demand between \acp{vm} $i$ and $j$ and assume that $d_{ij} = d_{ji}$.
We also consider a set, $A$, of $m$ clouds (i.e. \acp{ec}, or conventional clouds) that communicate over the Internet.
A \ac{vm} $i$ autonomously chooses its hosting cloud.

Let $x_i \in A$ denote the cloud that hosts $i$ and let $l(x_i, x_j) > 0$ be the average latency between $i$ and $j$ if they are hosted at $x_i$ and $x_j$ respectively (Note: if $i$ and $j$ are hosted at the same cloud $x_i = x_j$).
We assume that $l$ is reciprocal and monotonic.
Therefore, $l(x_i, x_j) = l(x_j, x_i)$ and there is an entirely nondecreasing order of $A \rightarrow A'$ such that for any consecutive $x_i, x_i' \in A'$, $l(x_i, x_j) \leq l(x_i', x_j)$.
The reciprocity condition ensures that measured latencies are aligned with peer-\acp{vm} and imitates the alignment rule in bird flocking.
We model $l(x_i, x_j) = \tau (x_i, x_j) + \rho (x_i) + \rho(x_j)$, where $\tau (x_i, x_j)$ is the average packet latency between $x_i$ and $x_j$, and $\tau (x_i, x_j) = \tau (x_j, x_i)$.
The quantity $\rho(x)$ is the average processing delay of $x$ modeled as: $\rho (x) = \delta \sum_{i \in V: x_i = x} \sum_{j\in V} d_{ij} / (\gamma(x) - \sum_{i \in V: x_i = x} \sum_{j\in V} d_{ij})$, where $\delta$ is an arbitrary delay constant and $\gamma(x)$ denotes the capacity of $x$ to handle all demanded traffic of its hosted \acp{vm}.
An increased value of $\rho(x_i)$ signals the \ac{vm} $i$ that it is crowding with other \acp{vm} in the same cloud, imitating the separation rule in bird flocking.

A \ac{vm} $i$ evaluates its weighted latency with its peers if hosted at $x$ as
\begin{equation}
  \label{eq:latency}
  u_i(x) = \sum_{j\in V} d_{ij} l(x, x_j) / \sum_{j\in V} d_{ij}.
\end{equation}
Our objective is to design an autonomous \ac{vm} migration protocol that converges to an outcome $\sigma=(x_1, x_2, \ldots, x_n)$ that minimizes the sum of weighted latency $\sum_{i \in V} u_i(x_i)$.
That is to say, $\sigma$ maximizes the responsiveness of all \acp{vm} given their peer-to-peer communication pattern (i.e. connectivity and demands).

\section{\protocol: Autonomous \ac{vm} Migration Protocol}
\label{proposed}

We design a simple protocol that allows a \ac{vm} to autonomously decide its hosting cloud among a set $A_i \subseteq A$ of available clouds by relying on only local regularized latency measurements.
%
%
We call $A_i$ the strategy set of $i$.
Every cloud $x$ evaluates its current weight $w_x = \sum_{i: x_i=x} u_i(x)$ and advertises a monotonic non-negative regularization function $f(w_x): \Rset^+ \rightarrow \Rset^+$, such that $\alpha < f(w_x) < 1$ for $\alpha > 0$, to all the \acp{vm} that are hosted at $x$.

Since each \ac{vm} is hosted by a single cloud and all \acp{vm} have access to the same strategy set, the VM migration problem is modelled as a singleton symmetric weighted congestion game with the objective of minimizing the social cost $C(\sigma) = \sum_{x \in A} w_x f(w_x)$.
If $f(w_x)$ is approximately $1$, this game model is approximately equivalent to minimizing $\sum_{i \in V} u_i(x_i)$.
Throughout, we use $\migrate{i}{x}{y}$ to mean that a \ac{vm} $i$ migrates from cloud $x$ to cloud $y$. Letting $\eta \leq 1$ be a design threshold, we propose the following migration protocol:
\vspace{-5pt}
\begin{algorithm}[H]
\centering
\begin{algorithmic}[1]
  \renewcommand{\algorithmicrequire}{\textit{Initialization:}}
  \renewcommand{\algorithmicensure}{\textit{Ensure:}}
  \REQUIRE Each \ac{vm} $i \in V$ runs at a cloud $x \in A$.
  \ENSURE  A Nash equilibrium outcome $\sigma$.
  \STATE During round $k$, do in parallel $\forall i \in V$
  \\ \textit{Greedy migration process imitating cohesion in flocking:}
  \STATE $i$ solicits its current strategy $A_i$ from $x$.
  \STATE $i$ randomly selects a target cloud $y \in A_i$.
  \IF { $u_i(y) f(w_y + u_i(y)) \leq  \eta  u_i(x) f(w_x - u_i(x))$}
    \STATE $\migrate{i}{x}{y}$.
  \ENDIF
\end{algorithmic}
\caption*{\protocol: Autonomous \ac{vm} migration protocol.}
\end{algorithm}


We now first begin by proving that \protocol~converges to a Nash equilibrium, where each \ac{vm} chooses a single cloud (strategy) and no \ac{vm} has an incentive (i.e. less average latency) to migrate from its current cloud.
Then, we derive an upper bound on \protocol's~\ac{poa} for general regularization functions, $f$, following a similar approach to \cite{bhawalkar2014weighted}.
Finally, we propose a regularization function $f(w_x) \approx 1 $ that achieves a tight \ac{poa} of at most $1+\epsilon$.

\comment{
Table~\ref{notation} summarizes key notation.

\begin{table}[!t]
\renewcommand{\arraystretch}{1.3}
\caption{Summary of key notation}
\label{notation}
\centering
\begin{tabular}{c||c}
\hline
\bfseries Symbol & \bfseries Definition\\
\hline\hline
$n$ & 2.0\\
$m$ & 2.0\\
$V$ & 2.0\\
$A$ & 2.0\\
$f$ & 2.0\\
$w_x$ & 2.0\\
$u_i$ & 2.0\\
$l(x_i, x_j)$ & 2.0 \\
$\sigma$ & 2.0 \\
\hline
\end{tabular}
\end{table}
}

\subsection{Equilibrium}

Convergence to a Nash equilibrium in \protocol~is non-obvious given the inter-dependency of a \ac{vm}'s average weight with its peers.

\begin{theorem}
  \protocol~converges to a Nash equilibrium outcome.
    \begin{proof}
    Any step of \protocol~corresponds to choosing a random outcome from a finite number of outcomes. We show that any step of \protocol~reduces the social value $C(\sigma)$ and $C(\sigma)$ is bounded below. We first show that if a \ac{vm} $i$ migrates from cloud $x$ to cloud $y$, the increase in $y$'s weight is less than the decease in $x$'s weight. We then use contraction to show that if a subset of $i$'s peers have a total latency that increased after $i$ migrates, the remaining peers must have a total latency that decreased by a greater value.
    Let $\sigma^{k}$ and $w_x^{k}$ denote the game outcome and the weight of $x$ at round $k$ respectively, and let
    $\Delta w_x = w_x^{k+1} - w_x^{k}$.
        If $\migrate{i}{x}{y}$, then
        $$ u_i(y) f(w_y + u_i(y)) \leq  \eta  u_i(x) f(w_x - u_i(x)).$$
        As $ \alpha < f < 1$, we have two extreme cases:  $ u_i(y) \alpha \leq  \eta  u_i(x) $, or $ u_i(y) \leq  \eta  \alpha u_i(x) $, which implies that:
        \begin{equation}
        \label{eq:state1}
          u_i(y) \leq \frac{\eta}{\alpha} u_i(x).
        \end{equation}
        Because of the migration:
        $\Delta w_x < 0$ and
        $\Delta w_y > 0$,
        but $\left| \Delta w_x \right| \geq  u_i(x)$,
        and $\left| \Delta w_y \right| \leq  u_i(y).$ otherwise $i$ would not migrate, then
        \begin{equation}
        \label{eq:state2}
          \left| \Delta w_y \right| \leq \left| \Delta w_x \right| .
        \end{equation}
        Let $z_i=u_i(x_i) f(w_{x_i})$ denote the regularized weighted latency and let $z_i^t$ denote its value at round $k$. After the migration, $i$'s peers split into two subsets:
        $V_{\text{inc}} = \{j \in V: z_j^{k+1} > z_j^{k} )\},$ and
        $V_{\text{dec}}= \{j \in V: z_j^{k+1} < z_j^{k} )\}.$
        Assume after the migration step that
        $\sum_{j \in V_{\text{inc}}} z_j  > \sum_{k \in V_{\text{dec}}} z_k.$
        For $f \approx 1$,
        $\sum_{j \in V_{\text{inc}}} u_j(x_j) > \sum_{k \in V_{\text{dec}}} u_k(x_k).$ Substitute with the weighted latency value from \eqref{eq:latency} and since the demands (i.e. $d_{ij}$) remain unchanged,
        $\sum_{j \in V_{\text{inc}}} l(x_j, y) > \sum_{k \in V_{\text{dec}}} l(x_k, y).$
        By the reciprocity of $l$,
        $\sum_{j \in V_{\text{inc}}} l(y, x_j) > \sum_{k \in V_{\text{dec}}} l(y, x_k),$ and
        $u_i(y) \geq u_i(x)$, which contradicts \eqref{eq:state1}.
        By this contradiction and from \eqref{eq:state2}, the social value $C(\sigma^{k+1}) \leq C(\sigma^{k})$. Since $n$ and $m$ are finite, the number of all possible outcomes is finite. An outcome after a \ac{vm} migration, $\sigma^{k+1}$, corresponds to randomly choosing an outcome from the finite outcome space which reduces the social value. Since $C(\sigma) > 0$, then \protocol~must converge to a Nash equilibrium. $\blacksquare$

    \end{proof}
\end{theorem}

\subsection{Price of Anarchy}

We first give a generic upper bound of the \ac{poa}, then we propose our regularization function that tightens the \ac{poa}.

\begin{lemma}
The social value of \protocol~has a perfect \ac{poa} at most $\lambda/ (1-\varepsilon)$ if for $\varepsilon < 1$ and $\lambda > 1 - \varepsilon$ the regularization function satisfies
$w^* f(w + w^*) \leq \lambda  w^* f(w^*) + \varepsilon w f(w) $, where $w \geq 0$ and $w^* > 0$.
\begin{proof} Let $\sigma$ denote a Nash outcome and $\sigma^*$ denote any alternative outcome. Also let $w_x$ and $w_x^*$ denote the weight on $x$ in outcomes $\sigma$  and $\sigma^*$ respectively. Similarly, let $x_i$ and $x_i^*$ denote $i$'s strategy (hosting cloud) in $\sigma$  and $\sigma^*$ respectively.
By definition of a Nash outcome, $\forall i,\; u_i(x_i) f(w_{x_i}) \leq u_i(x_i^*) f(w_{x_i^*}).$
Summing over all \acp{vm} we get,
$C(\sigma) = \sum_{i \in V} u_i(x_i) f(w_{x_i}) \leq \sum_{i \in V} u_i(x_i^*) f(w_{x_i^*})$
However,
\begin{equation*}
\begin{split}
\sum_{i \in V} u_i(x_i^*) f(w_{x_i^*}) &  \leq \sum_{x \in A}
\sum_{i: x_i = x} u_i(x) f(w_x + u_i(x) )  \\
 & \leq \sum_{x \in A}  w_x^* f(w_x + w_x^* ) \\
 & \leq \sum_{x \in A}  \lambda  w_x^* f(w_x^* ) + \varepsilon w_x f(w_x) \\
 &  = \lambda C(\sigma^*) + \varepsilon C(\sigma).
\end{split}
\end{equation*}
Then,  $C(\sigma) \leq \lambda C(\sigma^*) + \varepsilon C(\sigma).$
Rearranging we get,
$\text{POA} = C(\sigma)/C(\sigma^*) \leq \lambda / (1-\varepsilon).$
$\blacksquare$

\end{proof}
\end{lemma}

\begin{theorem}
The regularization function $f(w) = exp(-1/(w+a))$ tightens the \ac{poa} to $1+\epsilon$ for a sufficiently large constant $a$ and reduces the game to the original \ac{vm} migration problem, i.e. minimizing $\sum_{i} u_i(x_i)$.
\begin{proof}
For $$\lambda \leq \frac{f(w_{\textrm{max}} + w_{\textrm{min}})}{f(w_{\textrm{min}})} \left( 1- \varepsilon  \frac{w_{\textrm{max}} f(w_{\textrm{max}})}{w_{\textrm{min}} f(w_{\textrm{max}} + w_{\textrm{min}})} \right),$$
$f(w) = \exp(-1/(w+a))$ satisfies the condition $w^* f(w + w^*) \leq \lambda  w^* f(w^*) + \varepsilon w f(w)$ (verify by inspection).
For an infinitesimally small value of $\varepsilon$, we can choose $a$ such that $\lambda = 1+\epsilon$, hence the $\text{POA} \leq 1+\epsilon$. If $a$ is sufficiently large $f(w_x) \approx 1$, hence $C(\sigma) \approx \sum_{i} u_i(x_i)$. $\blacksquare$
\end{proof}
\end{theorem}

\section{Experimental Results and Discussion}

We numerically evaluate \protocol~using several simulation rounds. In each round we model the clouds as a complete graph with inter-cloud latency $\tau \sim \text{Uniform}(10, 100)$ and cloud capacity $\gamma \sim \text{Uniform}(50, 100)$. We simulate the peer-to-peer relations of \acp{vm} as a binomial graph with $d \sim \text{Uniform}(1, 10)$. Each simulated \ac{vm} runs \protocol~asynchronously.

\begin{figure}[htp]
  \centering
  \includegraphics[width=0.45\textwidth]{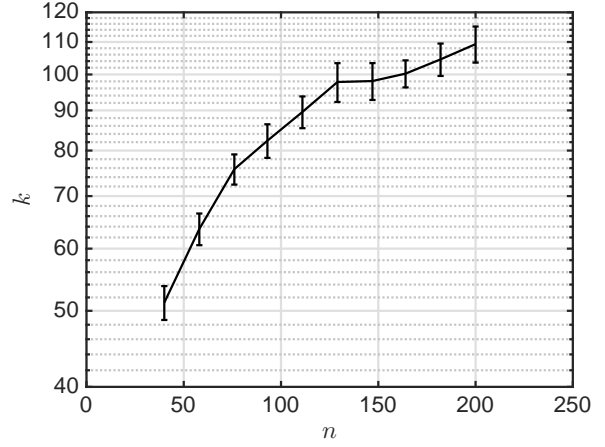}
  \caption{Convergence of \protocol: $m=37$, $\tau \sim \text{Uniform}(10, 100)$, $d \sim \text{Uniform}(1, 10)$, $a=9$, $\gamma \sim \text{Uniform}(50, 100)$, and $\eta=0.9$.}
  \label{convergence}
\end{figure}

\figurename~\ref{convergence} shows the average rounds, $k$, required to converge to a Nash equilibrium at $95\%$-confidence interval with $0.1$ error in simulations.
Although in the worst case $k=O(n \log (n f_{max}))$, where $f_{max}$ is the maximum value of the regularization function $f$ \cite{chien2007convergence}, \figurename~\ref{convergence} suggests that \protocol~scales better than $O(n)$ on average.
The proof of the average convergence time is complex and depends on properties of the social value $C(\sigma)$ and other parameters. We leave this proof for future work.

\begin{figure}[htp]
  \centering
  \includegraphics[width=0.45\textwidth]{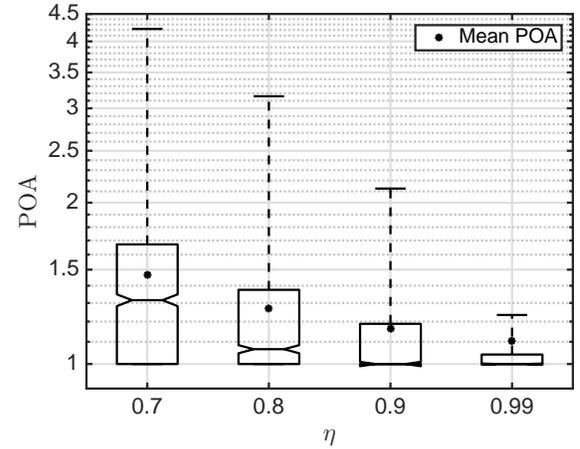}
  \caption{Price of Anarchy statistics for $m=5$, $n=8$, $\tau \sim \text{Uniform}(10, 100)$, $d \sim \text{Uniform}(1, 10)$, $\gamma \sim \text{Uniform}(50, 100)$, and $a=9$. }
  \label{fig:poa}
\end{figure}

\figurename~\ref{fig:poa} shows the \ac{poa} statistics of \protocol~with different $\eta$. We implemented the optimal solution as brute-force in simulations that evaluates the minimum social value among all possible \ac{vm} to cloud assignments. As $\eta \approx 1$, the maximum \ac{poa} achieved matched the theoretical value of $1.21$. In practice, it is desirable to to keep $\eta < 1$ (e.g. $\eta =0.7$) to avoid migrations with insignificant improvements and alternating migrations between clouds. Although the worst case \ac{poa} for $\eta = 0.7$ approaches $4.5$, the average \ac{poa} is acceptable in practice.

\subsection{Special Cases: Load-balancing and Energy Efficiency}
\label{special}

\protocol~can be redesigned with simple tweaks to suit other common cloud resource provisioning problems. We consider load-balancing and energy efficiency as special cases.

In load balancing, we seek an outcome $\sigma$ such that the total load is distributed proportionally to the clouds' capacities.
We first force the latency $\tau = 0$ between any cloud pairs $x,y \in A$. This is equivalent to marginalizing the effect of latency on the social value. We also force the \acp{vm} to ignore their peer relationships (i.e. $P=\emptyset$). The problem transforms immediately to minimizing the sum of cloud utilization. A \ac{vm} in this setup greedily migrates to the least loaded cloud.

\begin{figure}[htp]
  \centering
  \includegraphics[width=0.45\textwidth]{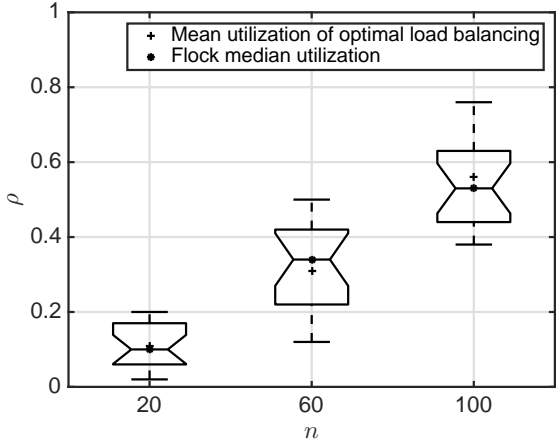}
  \caption{\protocol~usage for load balancing ($P=\emptyset$, $\tau = 0$) for $m=20$, $d \sim \text{Uniform}(1, 10)$, and $\gamma \sim \text{Uniform}(50, 100)$.}
  \label{fig:balance}
\end{figure}

\figurename~\ref{fig:balance} shows the ideal mean utilization for a load-balanced system ('+') and the mean utilization using \protocol~('*'). The box-plot also gives a summary statistics of the extreme value and the $75\%$-percentile samples. \protocol~performs very well as a load balancing protocol.

\begin{figure}[htp]
  \centering
  \includegraphics[width=0.45\textwidth]{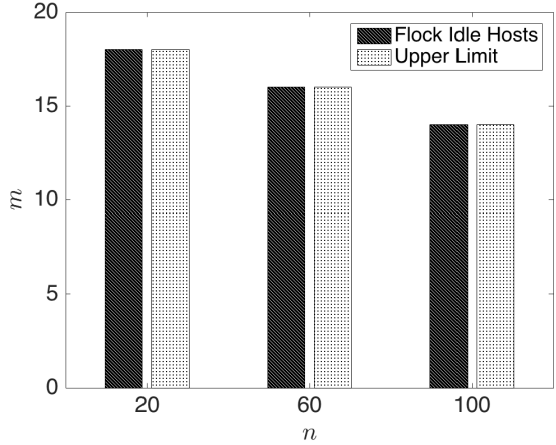}
  \caption{\protocol~usage for energy efficiency ($P= V \times V$, $\tau \rightarrow \infty$, $d = 1$) for $m=20$, and $\gamma \sim \text{Uniform}(50, 100)$.}
  \label{fig:energy}
\end{figure}

The energy-efficiency goal seeks an outcome $\sigma$ in which the maximum number of clouds are idle (i.e. with $\rho=0$). We (virtually) force each \ac{vm} to be connected to all other \acp{vm} (i.e. $P= V \times V$). We also force $\tau$ to be a very large value (i.e. $\tau \rightarrow \infty$, and $d = 1$). With this tweak, a \ac{vm} favors a cloud that hosts the largest number of \acp{vm} and with enough capacity $\rho$. \figurename~\ref{fig:energy} shows the ideal number of idle clouds that minimizes the energy consumption under a certain load (upper bound) and the number of idle clouds using \protocol~under the same load.

\subsection{Impact of system dynamics}
\label{dynamics}

We now study the impact of various system dynamics on \protocol's convergence. If we considered any two \protocol~steps at discrete times $(k-1)$ and $k$, the values of the \ac{vm}'s utilities, $u_i$, evolve randomly. Both external factors and \ac{vm} migrations can influence this evolution, which introduce difficulty in the convergence analysis of \protocol. The simplest approach to deal with such dynamics is to assume that changes in latencies, \ac{ec} capacities, and any other state are much slower than \protocol~convergence. This means that changes in $u_i$ due to external factors such as link bandwidths or \ac{ec} failures varies slowly. It also means that \protocol~steps are very rapid and \protocol~measures only average values of $l$ or any other system state. Fortunately, we have shown that \protocol~converges in a finite number of steps. However, we cannot guarantee that this number of steps are taken much faster than changes in any system state. We will follow the approach in \cite{kubrusly1973stochastic} to show that despite that \protocol~may take slow migration decisions, its convergence still holds if we modeled the network dynamics (e.g. $l$) as a continuous-time Markov process that has values that are generated by the migration steps of \protocol.

Consider modeling the utility of a \ac{vm} as a general discrete-time stochastic process such that:
$\forall k \in \integ, u_{i,k+1} = u_{i,k} + b_k \omega_{k}$, where $u_{i,k+1}$ is the utility value of \ac{vm} $i$ at the discrete-time $k$, $b_k$ is a design parameter, and $\omega_{i,k}$ is a random variable that takes values according to both external factors that impact $u$ and \protocol~migrations prior to time $k$. For all $k \in \integ$, the values of $\omega$ are determined by
$\omega_{i,k} = h(u_{i,k}, Y_{i,k})$ and $Y_{i,k}=\int_{k}^{k+1} f(m_i(t)) dt$, where $m_i(t)$ is a continuous-time Markov process of generator $G^{u_{i,k}}$ and with values in a finite set $\Mset, f: \Mset \leftarrow \real^K$ is an arbitrary mapping, and $h: \real^L \times \real^K \leftarrow \real^L$ is a bounded continuous Lipschitz function in $u$ and is uniformly distributed in $Y$. The Markov process $m_i(t)$ is irreducible and ergodic where $m_i(k) = \lim_{t \to k, t<k} m_i(t)$. We previously assumed that $u_{i,k}$ is bounded, which is true given the bounded latencies, \ac{vm} demands, and \ac{ec} capacities. We also can  enforce such bounded values of $u$ by projecting $u_{i,k}$ to a finite subset of $\real^L$. Finally, we assume that $b_k$ is positive and decreasing in $k$, such that $b_k$ is constant as $k \to \infty$, so that $\sum_k b_k = \infty$ and $\sum_k b_k^2 < \infty$.

By adopting the discrete time factor into the values of $u_{i,k}$, we transfer \protocol~into a class of stochastic approximation algorithms with controlled continuous-time Markov noise \cite{borkar2006stochastic}. As $b_k$ is a decreasing step size, the speed of variations in $u_{i,k}$ decreases and vanishes as $k$ increases. This is equivalent to modeling $m_i(t)$ as a Markov process with a fixed generator (e.g. when $\tau$ values are frozen), and converges to an ergodic behavior. Hence, as $k$ increases, \protocol~uses averaged values of $u_i$ and represents a stable dynamical system (see \cite{borkar2006stochastic} for formal proofs of the convergence of stochastic approximation algorithms). We call the modified migration protocol, controlled-\protocol~and is given as:
\vspace{-5pt}
\begin{algorithm}[H]
\centering
\begin{algorithmic}[1]
  \renewcommand{\algorithmicrequire}{\textit{Initialization:}}
  \renewcommand{\algorithmicensure}{\textit{Ensure:}}
  \REQUIRE Each \ac{vm} $i \in V$ runs at a cloud $x \in A$.
  \ENSURE  A Nash equilibrium outcome $\sigma$.
  \STATE During round $k$, do in parallel $\forall i \in V$
  \\ \textit{Greedy migration process imitating cohesion in flocking:}
  \STATE $i$ solicits its current strategy $A_i$ from $x$.
  \STATE $i$ randomly selects a target cloud $y \in A_i$.
  \IF { $u_{i,k}(y) f(w_y + u_{i,k}(y)) \leq  \eta  u_{i,k}(x) f(w_x - u_{i,k}(x))$}
    \STATE $\migrate{i}{x}{y}$.
    \STATE $x = y$
  \ENDIF
  \STATE Update $u_{i,k+1}(x) = u_{i,k}(x) + b_k f(w_x)$
\end{algorithmic}
\caption*{Controlled-\protocol: Autonomous \ac{vm} migration protocol.}
\end{algorithm}

In the above algorithm, $b_k$ is a decreasing step size and is left as a design parameter. For example $b_k=1/k$ satisfies the decreasing condition of $b_k$ and that $\sum_k b_k = \infty$ and $\sum_k a_k^2 < \infty$. If $b_k$ is constant, we would obtain weak convergence only. The function $f(w_x)$ serves as the random variable $\omega_{i,k}$ for a \ac{vm} $i$. In such case $f(w_x) \equiv h(u_{i,k}, Y_{i,k})$, where $Y_{i,k} = \int_{k}^{k+1} \sum_{j: x_j=x, j \neq i} u_{j,t}(x) dt$ and $u_{j,t}(x)$ is the continuous-time value of $u_j$. The Markov process $m_i(t)$ in this case represents the current outcome $\sigma^{k}$ and that $f(m_i(t)) = \sum_{j: x_j=x, j \neq i} u_{j,t}(x)$ is an arbitrary mapping that is bounded and continuous Lipschitz function. We can easily verify that the properties of stochastic approximation algorithms with controlled continuous-time Markov noise as described earlier are satisfied and convergence of Controlled-\protocol~is ensured given the various dynamics.

\subsection{Migration cost}
\label{migrationCost}

Migration of a \ac{vm} can be too costly to perform very frequently. Depending on the implementation details of a \ac{vm} and the workload it serves, migration can involve transferring a large data volume between \ac{ec}, which introduces a significant network overhead. For some cloud services, frequent migration can cause intolerable service interruption, as the ''down-time'' due to a migration can range from few milliseconds to seconds \cite{clark2005live}. Accounting for the migration cost in \protocol~can be desired for several applications. We incorporate the migration cost in \protocol~utility values by two methods.

First, we include the migration cost as an external dynamics in the \acp{vm}' utilities and use Controlled-\protocol~to minimize the utility, which also includes the average migration cost. Let $g_i(k) \in \{0,1\}$ indicate whether $i$ migrated at time $k$ or not. Also let $R_i(k+1) = \beta_i R_i(k) + (1-\beta_i) g_i(k)$ denote the average forgetting value of the migrations of \ac{vm} $i$, where $ 0 \leq \beta_i \leq 1$ is a \ac{vm} specific parameter that reflects the impact of frequent migrations on service disruptions. The migration cost, $C_i$, is an increasing function of $R_i(k)$, $C_i(R_i(k))$. To incorporate the migration cost, we redefine $u_i$ for the target cloud in \protocol~as $$u_i(y) = \sum_{j\in V} d_{ij} l'(y, x_j) / \sum_{j\in V} d_{ij},$$ where $l'(y, x_j)= l(y, x_j) + C_i(R_i(k)) + + C_j(R_j(k))$ and $u_i(x)$ (i.e. utility for current cloud) is memorized. This is equivalent to penalizing the latencies values measured with an additional dynamic that accounts for the estimated average number of migrations. As $R_i$ is a function in prior migration, the $C_i$ value increases as $i$ migrates more frequently and vanishes over time as $i$ pauses migrations. This tweak perceives the numerical properties of both $l$ and $u$, hence maintains the convergence and \ac{poa} results of Controlled-\protocol.

Although the first approach minimizes the average migration cost as $k \to \infty$, it requires exchanging the estimated migration cost between each $i, j \in E$ to maintain the reciprocity condition of $l$. This rapid message exchange introduces a significant communication overhead. Moreover, incorporating the average migration cost in $u_i$ only has long term benefit on the system as $k$ grows and it can causes undesirable short term service disruption for some applications. For such interruption-sensitive applications, it may be beneficial to delay the migration decisions using the $\eta$ parameter instead of incorporating the migration cost into the utility function. In this second approach each \ac{vm} evaluates its own $\eta_i$ accounting for the estimated number of migration as: $\eta_i(k) = \frac{1}{\exp (R(k))}$. As the \ac{vm}, $i$, performs less migrations over time, $\eta_i(k) \to 1$ and $i$ migrates for any slight improvement in its utility. Whereas if $i$ performs frequent migrations, $\eta_i(k) \to 0.36$, where $i$ only migrates if the migration decision brings a significant improvement to $u_i$. The drawback of this approach can be seen in \figurename~\ref{fig:poa}, where we sacrifice the tightness of the \ac{poa} as $\eta < 1$.

\section{Conclusion}
\label{conclusion}
We propose \protocol; a simple autonomous \ac{vm} migration protocol.
\protocol~considers the peer-to-peer interaction of \acp{vm} in heterogeneous edge and conventional cloud platforms.
We show that \protocol~converges to a Nash equilibrium with $(1+\epsilon)$ \ac{poa}.
\protocol~minimizes the average latency of \acp{vm} as a generic goal with diverse use cases and can be easily redesigned to serve other purposes such as load-balancing and energy efficiency.

\section{Acknowledgment}
This work was made possible by NPRP grant \# NPRP 5- 319-2-121 from the Qatar National Research Fund (a member of Qatar Foundation). The statements made herein are solely the responsibility of the authors.

The authors would like to thank Bassem Khalfi for his comments to make this paper more readable.

\bibliographystyle{IEEEtran}
\bibliography{IEEEabrv,./references}

\end{document}